\documentclass[aps,prb,twocolumn,showpacs,amsmath,amssymb,superscriptaddress,longbibliography]{revtex4-1}

\usepackage[english]{babel}
\usepackage{amsmath,amsfonts,bm,graphicx,verbatim,mathrsfs,units,color,nicefrac}
\usepackage[colorlinks=true,citecolor=blue]{hyperref}

\begin{document}
\title{Quantum theory of an electron waiting time clock}
\date{\today}

\author{David Dasenbrook}
\affiliation{D\'epartement de Physique Th\'eorique, Universit\'e de Gen\`eve, 1211
  Gen\`eve, Switzerland}
\author{Christian Flindt}
\affiliation{Department of Applied Physics, Aalto University,
  00076 Aalto, Finland}

\begin{abstract}
The electron waiting time is the time that passes between two subsequent charge transfers in an electronic conductor. Recently, theories of electron waiting times have been devised for quantum transport in Coulomb-blockade structures and for mesoscopic conductors, however, so far a proper description of a detector has been missing. Here we develop a quantum theory of a waiting time clock capable of measuring the distribution of waiting times between electrons above the Fermi sea in a mesoscopic conductor. The detector consists of a mesoscopic capacitor coupled to a quantum two-level system whose coherent precession we monitor. Under ideal operating conditions our waiting time clock recovers the results of earlier theories without a detector. We investigate possible deviations due to an imperfect waiting time clock. As specific applications we consider a quantum point contact with a constant voltage and lorentzian voltage pulses applied to an electrode.
\end{abstract}

\maketitle

\section{Introduction}

Recent breakthrough experiments have paved the way for giga-hertz quantum electronics.\cite{bocquillon14} Single electrons can now be emitted into low-dimensional circuits using either driven mesoscopic capacitors\cite{feve07} or by applying lorentzian voltage pulses to a contact.\cite{dubois13,jullien14} Along with these advances follows the need to characterize the accuracy of such dynamic single-electron sources. To this end one may investigate the electron waiting time.\cite{brandes08} This is the time that passes between two subsequent charge transfers in an electronic conductor. For an ideal single-electron emitter, the waiting time between subsequent emissions should be determined by the period of the external drive.\cite{albert11,dasenbrook14,albert14,dasenbrook15,hofer15} In reality, however, the waiting time fluctuates for instance due to the uncertainty in the emission time or because of cycle-missing events. These fluctuations can be characterized by the distribution of electron waiting times.

Theories of electron waiting times have been developed both for quantum transport in Coulomb-blockade structures\cite{brandes08,welack09,welack09b,albert11,thomas13,sothmann14,talbo15,brandes16} and for mesoscopic conductors.\cite{albert12,haack14,thomas14,dasenbrook14,dasenbrook15} Electron waiting times have also been investigated  in relation to transient quantum transport\cite{tang14,souto15} and for superconductors.\cite{rajabi13,dambach15,albert16} In some Coulomb-blockade structures, the tunneling of individual electrons can be monitored in real-time,\cite{fujisawa06,gustavsson06,gustavsson09,flindt09,ubbelohde12,maisi14} and the electron waiting time is clearly defined as the time that passes between two subsequent detections of a tunneling event. By contrast, in mesoscopic conductors, where the electronic transport is phase-coherent, the concept of electron waiting times is more subtle. In particular, it is not immediately obvious what physical process constitutes a detection event. As such, a proper definition of the electron waiting time relies on a careful description of a specific detector. In the context of full counting statistics, a quantum theory of a detector was developed by Levitov, Lee, and Lesovik.\cite{levitov96}

\begin{figure}
  \centering
  \includegraphics[width=.7\columnwidth]{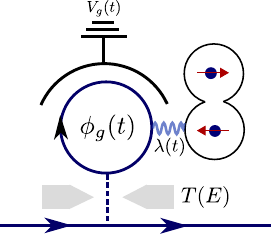}
  \caption{Electron waiting time clock. The clock consists of a mesoscopic capacitor coupled via a quantum point contact to a chiral edge state. Due to the energy-dependent transmission $T(E)$, only electrons above the Fermi level in the edge state can enter and leave the capacitor. Electrons inside the capacitor interact with a two-level system via the controllable coupling $\lambda(t)$. The gate potential $V_g(t)$ is used to empty the capacitor and leads to the time-dependent scattering phase $\phi_g(t)$. By monitoring the two-level system, the distribution of electron waiting times can be measured. The clock is placed after the scatterer whose WTD we wish to measure.}
  \label{fig:itpmeasurement}
\end{figure}

Existing theories of electron waiting times in mesoscopic conductors consider the electrons above the Fermi sea.\cite{albert12,haack14,thomas14,dasenbrook14,albert14,dasenbrook15,hofer15} For a fully transmitting single-channel conductor with an applied voltage $V$, it has been predicted that the distribution of electron waiting times should be given by a Wigner-Dyson distribution with the mean waiting time determined by the applied voltage as $\bar{\tau}=h/(eV)$.\cite{albert12,haack14} For typical voltages in the micro-volt regime, this mean waiting time is on the order of nano-seconds. This is a feasible time-scale from an experimental point of view. By contrast, if electrons in the Fermi sea are included, the mean waiting time would be given by the inverse Fermi energy, implying that a measurement of the electron waiting time essentially would be out of reach. Moreover, for dynamic single-electron sources, one is interested in the waiting time between the emission of electrons above the Fermi surface rather than in the intrinsic fluctuations in the Fermi sea.\cite{albert11,dasenbrook14,albert14,dasenbrook15,hofer15} For these reasons, theories of waiting times between electrons above the Fermi sea are attractive.

It is well-known that measurements of zero-frequency quantities like the average current and the shot noise only concern the electrons above the Fermi level.\cite{buttiker90scat,blanter00} On the other hand, measurements of the finite-frequency noise (and other short-time measurements) with standard current detectors are generally also sensitive to the underlying Fermi sea.\cite{gavish01} Bearing this in mind, it is clear that a theory of electron waiting times in mesoscopic conductors should include a description of a detector. This is the central goal of this paper. Specifically, we devise a quantum theory of a waiting time clock that is capable of measuring the distribution of waiting times between electrons above the Fermi sea in a mesoscopic conductor. When operated under ideal conditions, our waiting time clock recovers the results of earlier theories without a detector. Within our theoretical description, we can also investigate possible deviations due to imperfect operating conditions.

The rest of the paper is organized as follows. In Sec.~\ref{sec:fcswtd} we discuss the scattering theory of full counting statistics (FCS) in mesoscopic conductors with a specific emphasis on the detector. In Sec.~\ref{sec:wtd} we introduce the basic concepts of waiting time distributions (WTDs), including the idle time probability which will be important further on. In Sec.~\ref{sec:mescap} we describe our electron waiting time clock and its building blocks as indicated in Fig.~\ref{fig:itpmeasurement}. In Sec.~\ref{sec:applications} we illustrate the use of the waiting time clock with two specific applications. A possible implementation of the measurement scheme is described in Sec.~\ref{sec:fixedcouplings}. Finally, in Sec.~\ref{sec:conclusions}, we present our conclusions and give an outlook on future work.

\section{Time-resolved counting statistics}
\label{sec:fcswtd}
We start by recapitulating the scattering theory of time-dependent FCS with a special emphasis on the detector. An absorptive electron detector has been investigated theoretically in an early work.\cite{saito92} As an alternative, Levitov, Lee, and Lesovik later on considered a detector which conserves the number of electrons.\cite{levitov96} In this approach, the detector consists of a quantum two-level system, such as a spin-$\nicefrac[]{1}{2}$ particle, which rotates coherently in the magnetic field induced by the electrical current in the conductor. The moment generating function of the FCS can then be measured directly as a function of the coupling strength between the spin and the conductor.

To see this, we consider the combined system, including the detector, described by the Hamiltonian
\begin{equation}
  \label{eq:hamiltonian}
  \hat{H}(t) = \hat{H}_\text{el}(t) + \hat{H}_\text{int}(t) = \hat{H}_\text{el}(t) - \lambda(t)\frac{ \hbar}{2 e} \hat{\sigma}_z \hat{I}.
\end{equation}
The Hamiltonian of the conductor is denoted as $\hat{H}_\text{el}(t)$, $\hat{\sigma}_z$ is the Pauli matrix for the $z$-component of the spin, and $\hat{I}$ is the operator for the electrical current through a given cross-section of the conductor. The particular form of the coupling between the spin and the conductor makes the spin rotate in the $x$-$y$ plane of the Bloch sphere due to the magnetic field induced by the electrical current. The coupling strength $\lambda(t)$ is assumed to be controllable and generally time-dependent. We evolve the combined system from $t=-\infty$ to $t=\infty$ and describe the finite duration of a measurement by having a coupling which is only non-zero during the measurement. (In a different approach,\cite{muzykantskii03,schonhammer07} one detects the total charge in one of the leads at the beginning and at the end of the measurement and then defines the number of transferred charges as the difference between the two measurement outcomes.) After the complete time evolution, the electronic conductor is integrated out and the density matrix of the spin is obtained.

By evaluating the off-diagonal element of this reduced density matrix, one arrives at the
function\cite{belzig01,nazarov03,beaud13}
\begin{equation}
  \label{eq:mgfhamiltonian}
  \chi(\lambda) = \left \langle T \left\{e^{i \int_{-\infty}^\infty \mathrm{d} t \hat{H}_{-\lambda}(t)/ \hbar}
  \right\}\widetilde{T}\left\{ e^{-i \int_{-\infty}^\infty \mathrm{d} t \hat{H}_\lambda(t)/ \hbar}  \right\} \right \rangle,
\end{equation}
where $T$ and $\widetilde{T}$ denote time and anti-time ordering, respectively. The Hamiltonian
$\hat{H}_\lambda(t)$ is obtained from Eq.~\eqref{eq:hamiltonian} by replacing $\hat{\sigma}_z$ by
unity so that it only acts on the electronic degrees of freedom. The electronic conductor consists
of a central scatterer connected to electronic leads and is described by $\hat{H}_\text{el}(t)$. The
electrons are non-interacting so that a scattering problem can be formulated in terms of a
scattering matrix that we denote by~$\mathcal{S}$. To include the coupling to the spin, we solve the
scattering problem of the electrons interacting with the spin via the time-dependent coupling
$\lambda(t)$ and denote the resulting scattering matrix by $U_\lambda$. Since both $\lambda(t)$ and
$\hat{H}_\text{el}(t)$ can be time-dependent, neither $U_\lambda$ nor $\mathcal{S}$ are necessarily
diagonal in the energy representation. The combined scattering matrix is denoted as
$\mathcal{S}_\lambda$ and will be specified in more detail in the following sections.

Equation~\eqref{eq:mgfhamiltonian} can be evaluated by means of the Keldysh technique.\cite{kamenevbook} Specifically, it can be written as
\begin{equation}
  \label{eq:mgf}
  \chi(\lambda) = \det \left(1 - n_F \left[ 1 - \mathcal{S}_{-\lambda}^\dagger
      \mathcal{S}_\lambda \right] \right),
\end{equation}
which is known as the Levitov-Lesovik determinant formula. Here, $n_F$ is the occupation matrix of the leads and the involved matrices have indices both in the channel and energy spaces.

For the special case of a single-channel chiral system, e.~g.~a quantum Hall edge state, there is no channel index. However, due to the general time-dependence of the problem, $\mathcal{S}_\lambda $ is not diagonal in the energy representation. At zero temperature, $n_F$ is just a projector onto the filled states in the lead from which electrons enter the conductor. We can then write Eq.~\eqref{eq:mgf} as
\begin{equation}
  \label{eq:mgfchiral}
  \chi(\lambda) = \det \left( \mathcal{S}^\dagger_{-\lambda}
    \mathcal{S}_\lambda \right),
\end{equation}
where the matrix elements of $\mathcal{S}^\dagger_{-\lambda} \mathcal{S}_\lambda $ have been restricted to the initially filled states.

We now specify the interaction between the electrons and the spin. Due to the spin, electrons pick up the additional scattering phase
$\exp(i \lambda(t)/2)$. The scattering matrix $U_\lambda$ therefore has the matrix elements
\begin{equation}
  \label{eq:qubitelementstime}
  [U_\lambda]_{t,t'} = e^{i \lambda(t)/2} \delta(t-t')
\end{equation}
in the time representation. We take an abrupt switching,
\begin{equation}
  \label{eq:abruptcoupling}
  \lambda(t) = \lambda \Theta(t-t_0) \Theta(\tau-t+t_0),
\end{equation}
where $\Theta(t)$ is the Heaviside step function, $t_0$ is the starting point of the measurement, $\tau$ is the duration, and $\lambda$ is the coupling strength. In the energy representation, the matrix elements of $U_\lambda$ then become
\begin{equation}
  \label{eq:qubitelementsenergy}
  [U_\lambda]_{E,E'} = \delta(E-E') + K^\lambda_\tau(E-E'),
\end{equation}
having defined
\begin{equation}
  \label{eq:Klambdatau}
  K^\lambda_\tau(E) = \left(e^{i\lambda/2} - 1 \right) K_\tau(E)
\end{equation}
in terms of the sine kernel\cite{albert12,haack14,dasenbrook14}
\begin{equation}
  \label{eq:sinekernel}
  K_\tau(E) = \frac{2}{E} e^{-i \frac{E(\tau+2t_0)}{2 \hbar}} \sin\left( \frac{E\tau}{2
  \hbar}\right).
\end{equation}

If the reservoirs at $t=t_0$  are not in a superposition of different number eigenstates, Equation~\eqref{eq:mgf} can be interpreted as the moment generating function of the FCS.\cite{shelankov03} Specifically, from the inverse Fourier transformation
\begin{equation}
  \label{eq:fcs}
  P(n) = \frac{1}{2 \pi} \int_0^{2 \pi} \chi(\lambda) e^{i n \lambda} \mathrm{d} \lambda,
\end{equation}
we obtain the probability $P(n)$ that $n$ charges have passed through the conductor while the detector was on.  We note that backaction
effects due to the measurement device are fully included in this formalism.

It is instructive to consider the limit of long measurement times. In this case, we can take $\lambda(t) \equiv \lambda$ to be constant in Eq.~(\ref{eq:abruptcoupling}), such that $U_\lambda$ becomes diagonal in the energy representation. If, furthermore, the Hamiltonian $\hat{H}_\text{el}(t) = \hat{H}_\text{el}$ is not time-dependent, the determinant over energies in Eq.~\eqref{eq:mgf} reduces to a product,
\begin{equation}
  \label{eq:mgflongtimestationary}
  \chi(\lambda) = \prod_{E>0} \det \left(1 - [n_F]_{E,E} \left\{ 1 - [S_{-\lambda}^\dagger S_\lambda]_{E,E} \right\} \right),
\end{equation}
where the determinant is now taken only over the channel indices of the matrices. For a single-channel two-terminal conductor with the energy-dependent transmission probability $T(E)$, contributions from left and right moving electrons below the Fermi level cancel each other at zero temperature and only electrons above the Fermi level need to be included,
\begin{equation}
  \label{eq:mgflongtimeqpc}
  \chi(\lambda) = \prod_{E_F>E>E_F+eV} \left[(e^{i \lambda}-1) T(E) +1 \right].
\end{equation}
This type of FCS is known as generalized binomial statistics.\cite{hassler08,abanov08,abanov09} The result shows us that observables measured over a long time, for instance the mean current or any zero-frequency current correlator, are only affected by the electrons in the voltage window $[E_F,E_F+eV]$ above the Fermi level. In the following we will see that finite-time measurements are more involved as they may also be influenced by electrons below the Fermi level.

\section{Waiting time distributions}
\label{sec:wtd}

Building on the previous section, we are now ready to develop a quantum theory of an electron waiting time clock. We begin by establishing the general framework of WTDs before moving on to a detailed description of our detector.

In many physical systems, WTDs are used to characterize the random time that passes between two clicks of a detector. For example, in quantum optics, single-photon detectors can detect individual photons emitted from a light source.\cite{vyas88,carmichael89} Also in Coulomb-blockade quantum dots, a nearby conductor can be used to monitor the charge state of the quantum dots and thereby detect the tunneling of individual charges in real-time.\cite{fujisawa06,gustavsson06,gustavsson09,flindt09,ubbelohde12,maisi14} By contrast, in mesoscopic conductors, the detection of individual electrons is still challenging. Therefore, a careful analysis of the detection process is required.

Given a series of detection events, the waiting time is the time that elapses between two successive detections. The distribution of waiting times is denoted as $\mathcal{W} (\tau)$. For a stationary process, it can be related to the idle time probability $\Pi(\tau)$ as
\begin{equation}
  \label{eq:wtditp}
  \mathcal{W} (\tau) = \langle \tau \rangle \partial_\tau^2 \Pi(\tau),
\end{equation}
where $\langle \tau \rangle$ is the mean waiting time.\cite{albert12,haack14} The idle time probability is the probability that no detections occur during a time interval of length $\tau$, $[t_0,t_0+\tau]$. For stationary processes, the idle time probability is independent of $t_0$ due to the translational invariance in time.\cite{albert12,haack14} By contrast, for periodically driven systems, it is a two-time quantity $\Pi(t_0,\tau)$ which explicitly depends on $t_0$.\cite{dasenbrook14,dasenbrook15,hofer15} The idle time probability entering Eq.~(\ref{eq:wtditp}) is then obtained as an average over the period of the driving $\mathcal{T}$,\cite{dasenbrook14,dasenbrook15,hofer15}
\begin{equation}
  \label{eq:itpaveraged}
  \Pi(\tau) = \frac{1}{\mathcal{T}} \int_0^\mathcal{T} \Pi(t_0,\tau) \mathrm{d} t_0.
\end{equation}

The idle time probability can also be expressed in terms of the time-dependent FCS as
\begin{equation}
  \label{eq:itpfcs1}
  \Pi(\tau) = P(n=0,\tau)
\end{equation}
which is the $n=0$ component of the probability $P(n,\tau)$ to observe $n$ detections during a time interval of length~$\tau$. Alternatively, the idle time probability can be expressed in terms of the moment generating function of the FCS
\begin{equation}
  \label{eq:itpfcs2}
  \chi(\lambda,\tau) = \sum_{n=0}^\infty P(n,\tau)e^{in\lambda}.
\end{equation}
Specifically, we get $P(n=0,\tau)$ by an inverse Fourier transformation as
\begin{equation}
\label{eq:fouriertrans}
  P(n=0,\tau)=\frac{1}{2\pi}\int_0^{2\pi}\chi(\lambda,\tau) \mathrm{d} \lambda,
\end{equation}
or, since the number of detector clicks is non-negative, by formally taking the limit $\lambda \to i \infty$,
\begin{equation}
\label{eq:chi2infty}
  P(n=0,\tau)=\chi(\lambda \to i \infty,\tau).
\end{equation}
Experimentally, one may measure the moment generating function $\chi(\lambda,\tau)$ for different values of the coupling strength $\lambda$ and from those measurements evaluate the idle time probability using the Fourier transformation in Eq.~(\ref{eq:fouriertrans}). On the theory side, it will be useful rather to take the limit $\lambda \to i \infty$ according to Eq.~(\ref{eq:chi2infty}). This also holds at finite temperatures, if the detector can only produce a non-negative number of clicks.

The considerations above rely on a detector that produces a series of clicks. Since single-electron detection remains challenging in mesoscopic conductors, we proceed here along a different route and instead develop a detector than can measure the idle time probability of electrons above the Fermi sea in a mesoscopic conductor. As we will see, this leads to a well-defined distribution of waiting times between subsequent electron transfers.

\section{Electron waiting time clock}
\label{sec:mescap}

The electron waiting time clock is depicted in Fig.~\ref{fig:itpmeasurement}. It consists of a mesoscopic capacitor\cite{buttiker93,pretre96} coupled to a two-level quantum system, such as a spin-$\nicefrac[]{1}{2}$ particle, in a similar spirit to the proposal to measure FCS by Levitov, Lee, and Lesovik.\cite{levitov96} The coupling  $\lambda(t)$ between the spin and the capacitor is controllable and time-dependent. We assume that the capacitor is initially depleted of electrons. As we will see, this setup makes it possible to measure the idle time probability and thus the WTD of electrons above the Fermi level in the incoming channel.

We start by constructing the scattering matrix of the electron waiting time clock. The capacitor is implemented with chiral edge states in the quantum Hall regime.\cite{feve07} Incoming electrons in the edge state on the left may be transmitted into the capacitor via a QPC and make one or several round trips inside the capacitor before leaving via the outgoing edge state to the right. While being inside the capacitor, the electrons interact with the spin that we monitor. Importantly, as we discuss below, we use a QPC with a cut-off in the transmission close the the Fermi level.

The scattering matrix of the waiting time clock is obtained by summing up the amplitudes of all possible scattering processes. Formally, we can express it as
\begin{equation}
  \label{eq:S_lambda}
  \mathcal{S}_\lambda= \mathcal{P}_R - \mathcal{P}_T  \left[\mathcal{S}^{(l)}_\lambda \sum_{n=0}^\infty \left(\mathcal{P}_R  \mathcal{S}^{(l)}_\lambda\right)^n\right] \mathcal{P}_T,
\end{equation}
where the first term describes processes where electrons are reflected on the QPC and never enter the capacitor. The second term describes processes where electrons enter the capacitor and complete $n+1$ round trips (or loops) inside the capacitor before leaving via the outgoing edge state. We now specify each matrix in this expression.

We consider a QPC with an energy-dependent transmission $T(E)$. In a strong magnetic field, the transmission takes the form\cite{buttiker90}
\begin{equation}
  \label{eq:qpctransmissionfunction}
  T(E) = \frac{1}{e^{\mathcal{B}(E_F-E)}+1},
\end{equation}
where the parameter $\mathcal{B}$ can be controlled by the magnetic field. The QPC is tuned such that the transmission is cut off at the Fermi energy $E_F$. For a sharp cut-off, only electrons above the Fermi level are allowed to enter and leave the capacitor. For a smooth cut-off, the measurement may be affected by electrons in the Fermi sea as we discuss in Sec.~\ref{sec:smooth}. The corresponding transmission and reflection matrices in Eq.~(\ref{eq:S_lambda}) read
\begin{equation}
  \label{eq:qpcprojectors}
  \begin{split}
  [\mathcal{P}_T]_{E,E'} &= \sqrt{T(E)}\delta(E-E') ,\\
  [\mathcal{P}_R]_{E,E'} &= \sqrt{1-T(E)}\delta(E-E').
  \end{split}
\end{equation}

Next, we define the scattering matrix $\mathcal{S}^{(l)}_\lambda$ describing one round trip inside the capacitor. An electron inside the capacitor can make one or several round trips. For each completed loop, it picks up the scattering phase
\begin{equation}
[\mathcal{S}^{(l)}_\lambda]_{t,t'}=e^{i (\phi_g(t)+\lambda(t) / 2)}\delta(t-t'-\tau_0),
\end{equation}
where $\tau_0=\ell/v_F$ is the time it takes to complete one loop with $\ell$ being the circumference of the capacitor and $v_F$ the Fermi velocity. The specific times when the electron enters and leaves the capacitor are denoted as $t'$ and $t$, respectively. The phase $\phi_g(t)$ picked up during one loop due to the time-dependent gate-voltage $V_g(t)$ reads
\begin{equation}
\phi_g(t)=\frac{e}{\hbar}\int_{t-\tau_0}^t V_g(t')\mathrm{d} t'.
\end{equation}
As we will see below, it is convenient to apply a linearly rising gate voltage of the form
\begin{equation}
\label{eq:voltage}
V_g(t)=\delta V_g (t/\tau_0+1/2),
\end{equation}
where $\delta V_g$ is the increase of the voltage during one loop. In this case, the phase takes the simple form
\begin{equation}
\phi_g(t)=\frac{e\delta V_g}{\hbar} t.
\end{equation}
Finally, the coupling to the spin $\lambda(t)$ is given by Eq.~(\ref{eq:abruptcoupling}).

In the energy representation, the scattering matrix is non-diagonal with matrix elements reading
\begin{equation}
\label{eq:Sloopmatrixelements}
\begin{split}
  [\mathcal{S}^{(l)}_\lambda]_{E,E'} = &\left[ \delta(E´-E'-e\delta V_g) + K_\tau^\lambda(E-E'-e\delta V_g)
    \right]\\
    &\times e^{i (E'+e\delta V_g) \tau_0 / \hbar}.
\end{split}
\end{equation}
This is the probability amplitude for a particle with incoming energy $E'$ to change its energy to $E$ due to the interaction with the spin and the time-dependent voltage.

Having specified the various scattering matrices, we can construct the scattering matrix of the electron waiting time clock according to Eq.~(\ref{eq:S_lambda}). Moreover, if an additional scatterer (whose WTD we wish to measure) with scattering matrix $\mathcal{S}_\text{sys}$ is placed before the waiting time clock, the full scattering matrix becomes
\begin{equation}
  \label{eq:totalSmatrix}
  \mathcal{S}_\lambda^{(\text{tot})} = \mathcal{S}_\lambda \mathcal{S}_\text{sys}.
\end{equation}
In the following section, where we apply our method, we specify $\mathcal{S}_\text{sys}$ for two particular scatterers.

We start by considering the limit of a sharp cut-off in Eq.~(\ref{eq:qpctransmissionfunction}), where ${\mathcal{B}\gg 1/E}$ for all relevant energies. In this case, only electrons above the Fermi level are allowed to enter and leave the capacitor. Mathematically, the transmission and reflection matrices in Eq.~(\ref{eq:qpcprojectors}) become projectors onto energies above and below the Fermi level which we denote as $P_T$ and $P_R$, respectively. We can then evaluate the geometric series in Eq.~\eqref{eq:S_lambda} and write the scattering matrix as
\begin{equation}
  \label{eq:Smatrix}
  \begin{split}
  \mathcal{S}_\lambda &= P_R - P_T \mathcal{S}^{(l)}_\lambda P_T \\
  &- P_T \mathcal{S}^{(l)}_\lambda P_R(1-P_R\mathcal{S}^{(l)}_\lambda P_R)^{-1}P_R \mathcal{S}^{(l)}_\lambda P_T,
   \end{split}
\end{equation}
having used properties of the projectors. This expression has a clear physical interpretation as we now discuss.

The first term corresponds to electrons below the Fermi level which are reflected on the QPC and never enter the capacitor. The second term describes electrons above the Fermi level that enter the capacitor, interact with the spin and the time-dependent voltage, but stay above the Fermi level, so that they leave the capacitor after having completed just one loop. The third term describes electrons that complete more than one loop. Read from right to left, this term corresponds to processes, where an electron above the Fermi level enters the capacitor and is scattered below the Fermi level during the first loop as described by the matrix product $P_R \mathcal{S}^{(l)}_\lambda P_T$. The electron then completes a number of loops (possibly none) below the Fermi level. This is described by the matrix inversion $(1-P_R\mathcal{S}^{(l)}_\lambda P_R)^{-1}$, which can be re-expanded as a geometric series. Finally, in one last loop, the electron is scattered back above the Fermi level and leaves the capacitor as described by the matrix product $P_T \mathcal{S}^{(l)}_\lambda P_R$.

Ideally, the electron waiting time clock would be described by only the two first terms of the scattering matrix in Eq.~(\ref{eq:Smatrix}). Electrons above the Fermi level then interact only once with the spin, while electrons below the Fermi level are filtered out. However, due to the time-dependence of the measurement procedure, the third term is generally present. To suppress processes where electrons complete several loops and interact with the spin more than once, we apply the top-gate voltage. With a sufficiently large voltage increase, we can ensure that essentially all electrons end up with an energy above the Fermi level after having completed the first loop, even if they may have lost energy by interacting with the spin. They will then leave the capacitor via the QPC after having interacted with the spin only once.

\begin{figure*}
  \centering
  \includegraphics[width=\columnwidth]{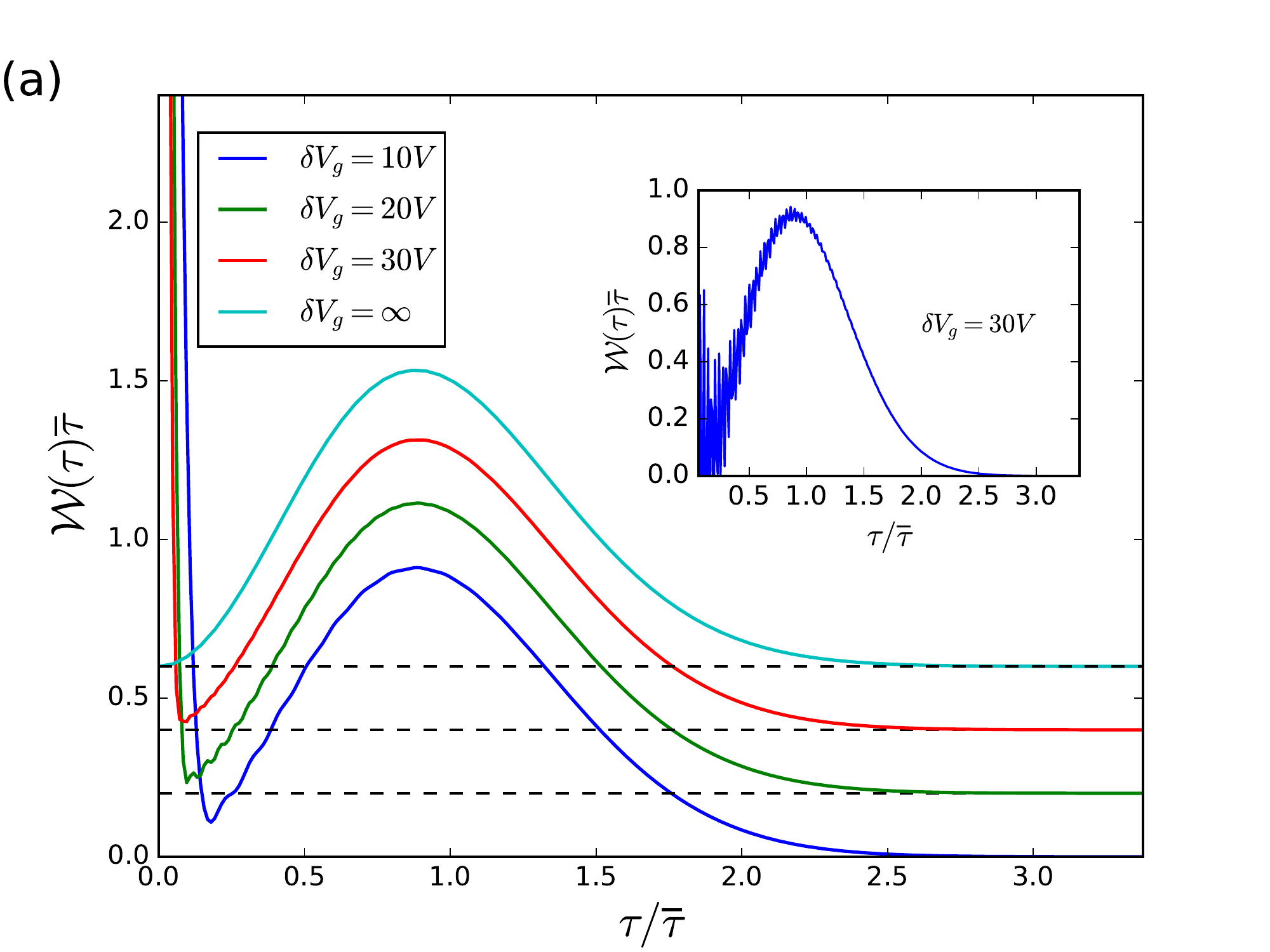}
  \includegraphics[width=\columnwidth]{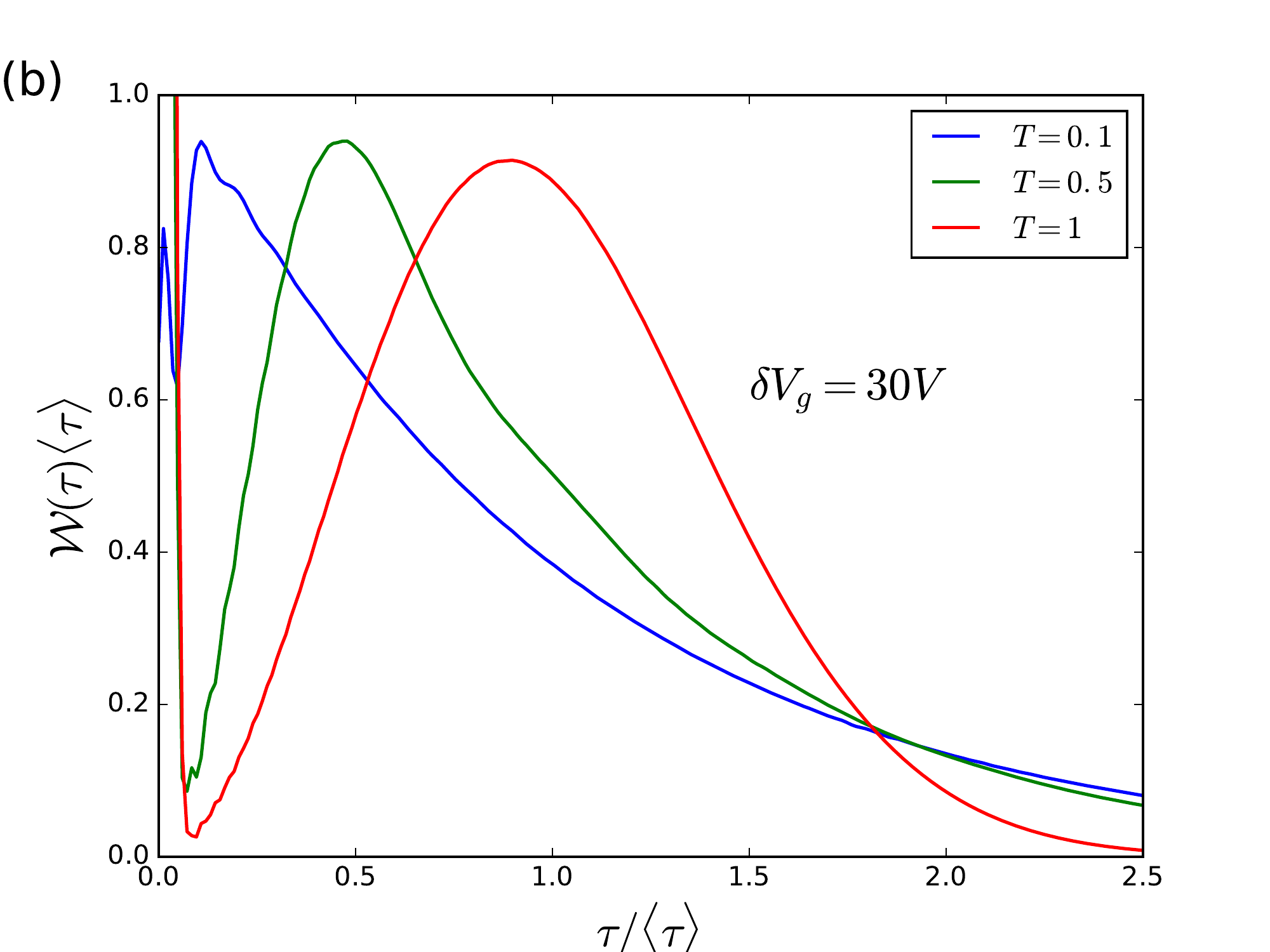}
  \caption{WTDs for a voltage-biased QPC. (a) Distribution of waiting times for a fully transmitting QPC ($T=1$) with an applied voltage $V$. The mean waiting time is $\bar{{\tau}}=h/(eV)$. We show results for different values of the gate voltage increase $\delta V_g$. With $\delta V_g=30V$, we essentially recover the prediction $(\delta V_g=\infty)$ of earlier theories without a detector.\cite{albert12,haack14} The curves have been shifted vertically by multiples of $0.2$ for the sake of a better visibility. We have applied a low pass filter to remove high-frequency oscillations. The inset shows the WTD for $\delta V_g= 30 V$ without the low pass filter. (b) Distribution of waiting times for different values of the QPC transmission $T$. We find a crossover from Wigner-Dyson distribution at full transmission ($T=1$) to Poisson statistics close to pinch-off ($T=0.1$) in agreement with earlier work without a detector.\cite{albert12,haack14}}
  \label{fig:wtd}
\end{figure*}

In this case, we can write the scattering matrix as
\begin{equation}
  \label{eq:Slambdaonlypositive}
  \mathcal{S}_\lambda = P_R - P_T \mathcal{S}_\lambda^{(l)} P_T,
\end{equation}
without processes involving several loops. In the following section, we discuss the values of $\delta V_g$ in Eq.~(\ref{eq:voltage}) needed for this to be a good approximation. To evaluate the determinant formula in Eq.~(\ref{eq:mgf}), we first note that
\begin{equation}
  \label{eq:SmltotSltot}
  (\mathcal{S}_{-\lambda}^{(\text{tot})})^\dagger \mathcal{S}_\lambda^{(\text{tot})} =
  \mathcal{S}_\text{sys}^\dagger \left(P_R +
  P_T (\mathcal{S}_{-\lambda}^{(l)})^\dagger    \mathcal{S}_\lambda^{(l)} P_T\right)
  \mathcal{S}_\text{sys},
\end{equation}
having used $[P_T, \mathcal{S}_\lambda^{(l)}]=0$, since electrons that enter the capacitor remain above the Fermi level. This holds for large values of $\delta V_g$. A simple calculation now shows that
\begin{equation}
(\mathcal{S}_{-\lambda}^{(l)})^\dagger    \mathcal{S}_\lambda^{(l)}=1+(e^{i\lambda}-1)\mathcal{K}_\tau,
\end{equation}
with the matrix elements
\begin{equation}
[\mathcal{K}_\tau]_{E,E'}=K_\tau(E-E')
\end{equation}
given by the sine kernel in Eq.~(\ref{eq:sinekernel}). Inserting these expressions into Eq.~(\ref{eq:mgfchiral}) we find at zero temperature
\begin{equation}
  \label{eq:mgfonlypositive}
  \chi(\lambda) = \det \left( 1 + \left(e^{i\lambda} - 1 \right)
  \mathcal{S}_\text{sys}^\dagger P_T \mathcal{K}_\tau    P_T \mathcal{S}_\text{sys} \right),
\end{equation}
having used that the scattering matrix $\mathcal{S}_\text{sys}$ is unitary and $P_T+P_R=1$. In this expression, the increase of the gate voltage $\delta V_g$ has dropped out. Moreover, by introducing the matrix
\begin{equation}
\label{eq:Qmatrix}
 \mathcal{Q}_\tau=  \mathcal{S}_\text{sys}^\dagger P_T \mathcal{K}_\tau    P_T \mathcal{S}_\text{sys},
\end{equation}
we may express the moment generating function as
\begin{equation}
 \chi(\lambda)=  \det \left( 1 + \left(e^{i\lambda} - 1 \right)  \mathcal{Q}_\tau \right).
 \end{equation}
Finally,  by taking the limit $\lambda \to i \infty$, we recover the determinant formula for the idle time probability of electrons above the Fermi level
\begin{equation}
  \label{eq:itponlypositive}
  \Pi(\tau) = \det \left( 1 -  \mathcal{Q}_\tau \right)
\end{equation}
previously derived in Refs.~\onlinecite{albert12,haack14,dasenbrook14} without specifying a detector. For a static scatterer with an applied voltage $V$, the matrix $\mathcal{Q}_\tau$ only has non-zero elements in the transport window $[E_F,E_F+eV]$.\cite{albert12,haack14}

Next, we consider the non-ideal situation where electrons may complete several loops inside the capacitor and interact with the spin more than once. This is described by the scattering matrix in Eq.~(\ref{eq:Smatrix}). We then evaluate the idle time probability by inserting this scattering matrix into Eq.~(\ref{eq:mgfchiral}). In contrast to Eq.~\eqref{eq:mgfonlypositive}, the function $\chi(\lambda)$ now contains terms that are proportional to $\exp(i\lambda / 2)$ as shown in App.~\ref{sec:appA}. Consequently, the function is not $2\pi$-periodic in $\lambda$ as required for a moment generating function according to Eq.~(\ref{eq:itpfcs2}), and the transport process cannot be described by a time series of discrete detection events. This is not a consequence of measurement back action in particular, but may possibly be related to the occurrence of negative probabilities in FCS due to interference effects.\cite{hofer15negative} Still, we can take the limit $\lambda \to i \infty$ and calculate the resulting WTD using Eq.~(\ref{eq:wtditp}). However, as we will see, the WTD may then become negative for certain waiting times. This is due to processes where an electron interacts more than once with the spin and thereby tampers with the measurement of the idle time probability.

\section{Applications}
\label{sec:applications}

We are now ready to illustrate the electron waiting time clock with two specific applications: A voltage-biased QPC and lorentzian voltage pulses. We also investigate the influence of a smooth transmission profile. In App.~\ref{sec:lorentzianswitching} we consider a smooth coupling to the spin instead of the abrupt switching given by Eq.~(\ref{eq:abruptcoupling}). Technically, it is worth mentioning that the Fredholm determinants that appear for example in Eq.~\eqref{eq:mgfchiral} can be evaluated efficiently using the algorithm described in Ref.~\onlinecite{bornemann10}.

\subsection{Voltage-biased QPC}

We start by considering a QPC with transmission probability $T$ and applied voltage $V$. In this case, we have $\mathcal{S}_\text{sys}=\sqrt{T  }$. Earlier works\cite{albert12,haack14} without a detector have shown that the WTD should display a cross-over from Wigner-Dyson statistics at full transmission ($T=1$) to a Poisson distribution close to pinch-off ($T\simeq 0$) with the mean waiting time given as
\begin{equation}
\label{eq:qpcmeanwait}
\langle\tau\rangle = \frac{\bar{\tau}}{T},
\end{equation}
where
\begin{equation}
\bar{\tau} = \frac{h}{eV},
\end{equation}
is the meaning waiting time at full transmission.

\begin{figure*}
  \centering
  \includegraphics[width=\columnwidth]{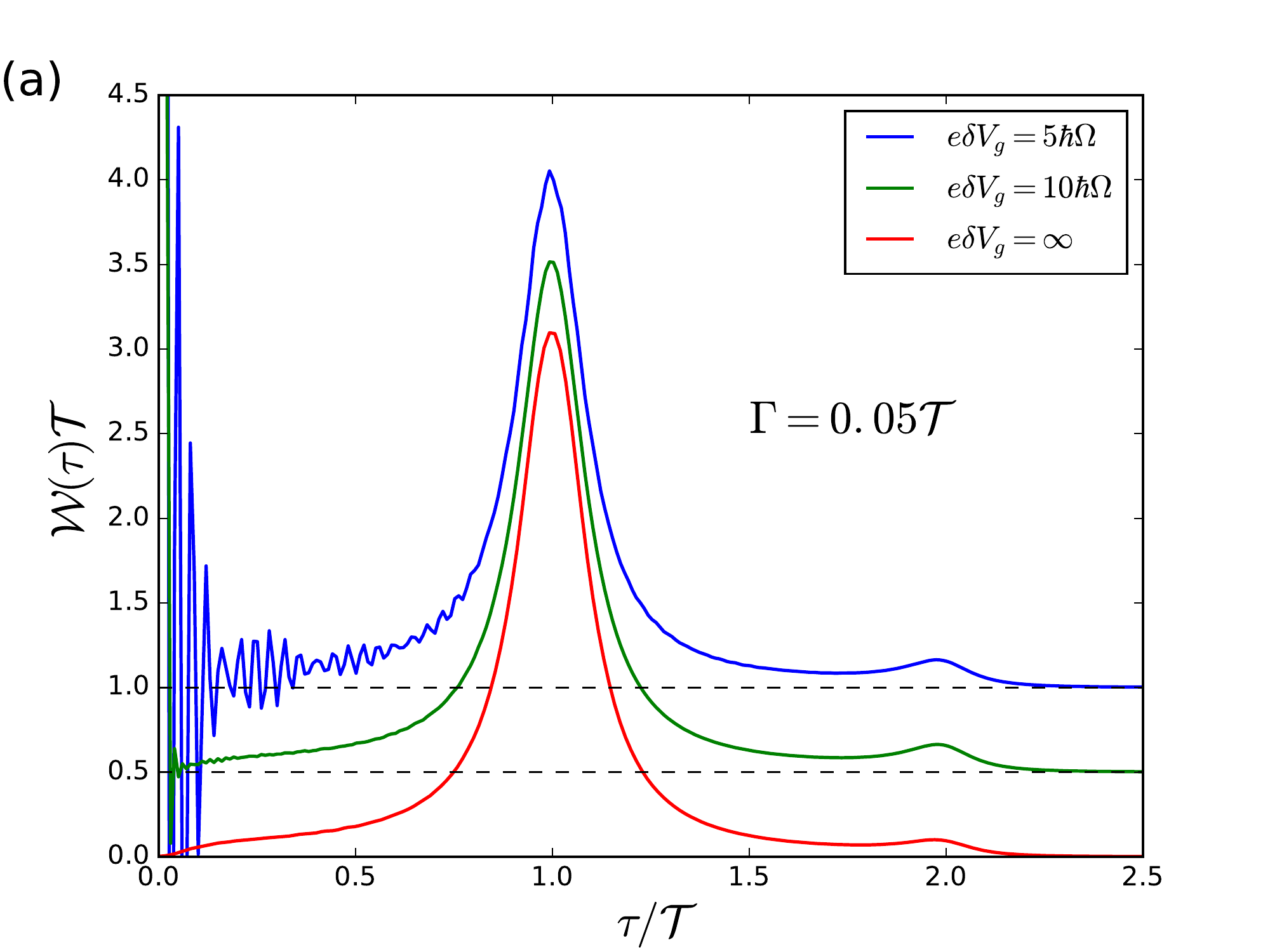}
  \includegraphics[width=\columnwidth]{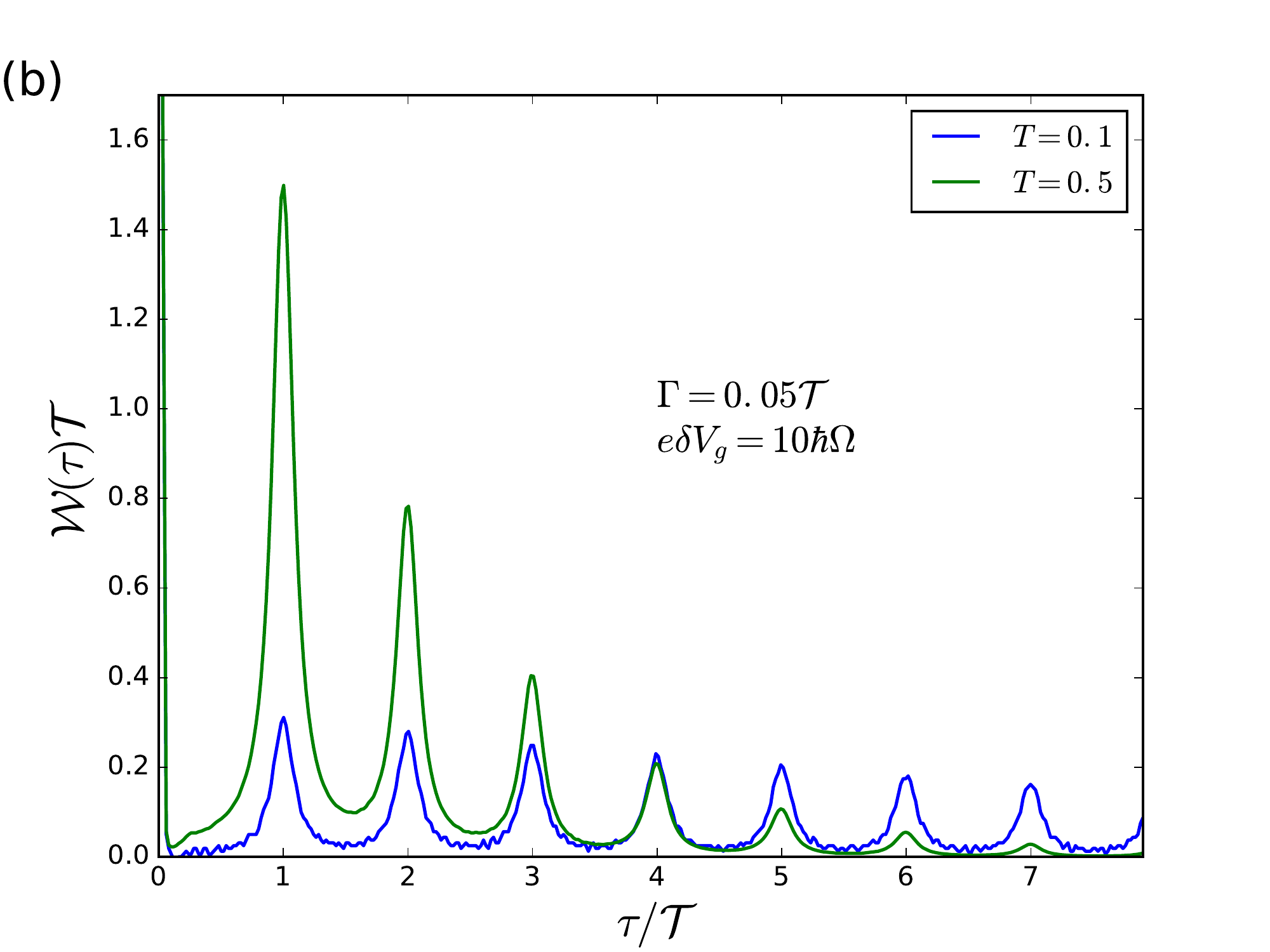}
  \caption{WTDs for levitons transmitted through a QPC. (a) Results for full transmission ($T=1$) with different values of the gate voltage increase $\delta V_g$. Already with $e\delta V_g =10 \hbar \Omega$, the waiting time clock reproduces results $(\delta V_g=\infty)$ of earlier theories without a detector.\cite{dasenbrook14,albert14} The curves have been shifted vertically for the sake of a better visibility. (b) Results for a QPC with a finite transmission $T$. In this case, levitons may reflect back on the QPC, and the WTD has peaks at multiplies of the period. }
  \label{fig:levitons}
\end{figure*}

Figure~\ref{fig:wtd} shows WTDs obtained with the waiting time clock. We calculate the idle probability using Eqs.~(\ref{eq:mgfchiral},\ref{eq:Smatrix}), see also App.~\ref{sec:appA}, and differentiate it twice with respect to $\tau$ according to Eq.~(\ref{eq:wtditp}). In panel (a) we show the WTD for a fully open QPC with different increments of the gate voltage $\delta V_g$. To measure the WTD, the coupling to the spin is only non-zero during a period of time on the order of $\bar{\tau}$. Such short detector pulses can change the energy of an electron by an amount on the order of $h/\bar{\tau}=eV$. Thus, to ensure that no electrons are scattered below the Fermi level during the measurement, the increase of the gate voltage must be much larger than this energy scale, i.~e. $\delta V_g \gg V$. This physical picture is confirmed by panel (a). As we increase $\delta V_g$, we approach the results of an ideal clock obtained from Eq.~(\ref{eq:itponlypositive}).

The time resolution of the waiting time clock depends on $\delta V_g$. A finite value of $\delta V_g$ introduces fluctuations in the WTD on the time scale $h/(e\delta V_g)$. The fluctuations essentially disappear for waiting times that are longer than the mean waiting time, where the measurement-induced disturbances are almost negligible. By contrast, for very short waiting times, the WTD may become negative as seen in the inset. To remove any spurious fluctuations, we apply a  low pass filter that suppresses frequencies on the order of $e\delta V_g / h$. The inset of panel (a) shows the WTD without the low pass filter.

In panel (b) we consider the WTD for different values of the QPC transmission $T$. The figure illustrates how our electron waiting time clock allows a observation of the cross-over from Wigner-Dyson distribution at full transmission to Poisson statistics close to pinch-off as previously predicted by theories without a detector.\cite{albert12,haack14}

\subsection{Lorentzian voltage pulses}

Next, we consider lorentzian voltage pulses applied to the input lead.\cite{dubois13,jullien14,levitov96,keeling06,ivanov97,lebedev05} The applied voltage has the form
\begin{equation}
  \label{eq:lorentzianpulses}
  V(t) = \sum_{j=-\infty}^\infty \frac{2 \hbar \Gamma}{(t-j \mathcal{T})^2 + \Gamma^2},
\end{equation}
where $\Gamma$ denotes the pulse width and $\mathcal{T}$ is the period. The voltage can be encoded in a time-dependent scattering phase picked up by electrons as they leave the lead
\begin{equation}
  \label{eq:timedepphase}
  e^{i \phi(t)} = e^{-i \frac{e}{\hbar} \int_{-\infty}^t  V(t')\mathrm{d} t'}
\end{equation}
By Fourier transforming this scattering phase, we obtain a Floquet scattering matrix with elements\cite{dasenbrook14,dubois13prb}
\begin{equation}
  \label{eq:levitonsfloquetsmatrix}
  \mathcal{S}_F(E_n,E) = \begin{cases}
    -2 e^{-n \Omega \Gamma} \sinh(\Omega \Gamma) & \text{if } n>0 \\
    e^{-\Omega \Gamma} & \text{if } n=0 \\
    0 & \text{otherwise}
  \end{cases},
\end{equation}
where $\Omega=2\pi/\mathcal{T}$ and $E_n=E+n \hbar \Omega$ with $n$ being an integer. For a periodic voltage, electrons can in general emit or absorb energy quanta of size $\hbar\Omega$. However, for lorentzian pulses (and only in this case), electrons can only absorb energy. Moreover, each pulse excites just a single electron-hole pair out of the Fermi sea without creating any additional disturbances.\cite{Note1} These single-electron excitations are known as levitons.\cite{dubois13,jullien14} The corresponding scattering matrix reads\cite{moskalets11book}
\begin{equation}
  \label{eq:Ssysperiodic}
  [\mathcal{S}_\text{sys}]_{E,E'} =  \sqrt{T}\sum_n \delta(E'-E_n) \mathcal{S}_F(E,E'),
\end{equation}
having included a QPC that reflects a fraction $R=1-T$ of the levitons before they reach the waiting time clock.

Figure~\ref{fig:levitons} shows the distribution of waiting times between levitons measured with the waiting time clock. For a fully transmitting QPC, the WTD is peaked around the period of the driving $\mathcal{T}$, panel~(a). Unlike the results for the voltage-biased QPC, there is no need to apply a low pass filter. We still observe small oscillations with a period of $h/e\delta V_g$, but they essentially disappear already for $\delta V_g =10 \hbar \Omega/e$. Physically, the levitons are well-localized in time and space, and one would expect that they are easier to distinguish from the underlying Fermi sea than electrons emitted from a constant voltage source. This is indeed confirmed by our results.

In panel (b) we consider the WTD of levitons transmitted through a partially reflecting QPC. In this case, levitons may reflect back on the QPC. As a consequence, the WTD develops peaks at multiplies of the period, with each peak corresponding to the number of subsequent reflections that have occurred. Again, we find good agreement with earlier theories without a detector.\cite{dasenbrook14,albert14}

\subsection{Smooth QPC transmission}
\label{sec:smooth}
\begin{figure}
  \centering
  \includegraphics[width=\columnwidth]{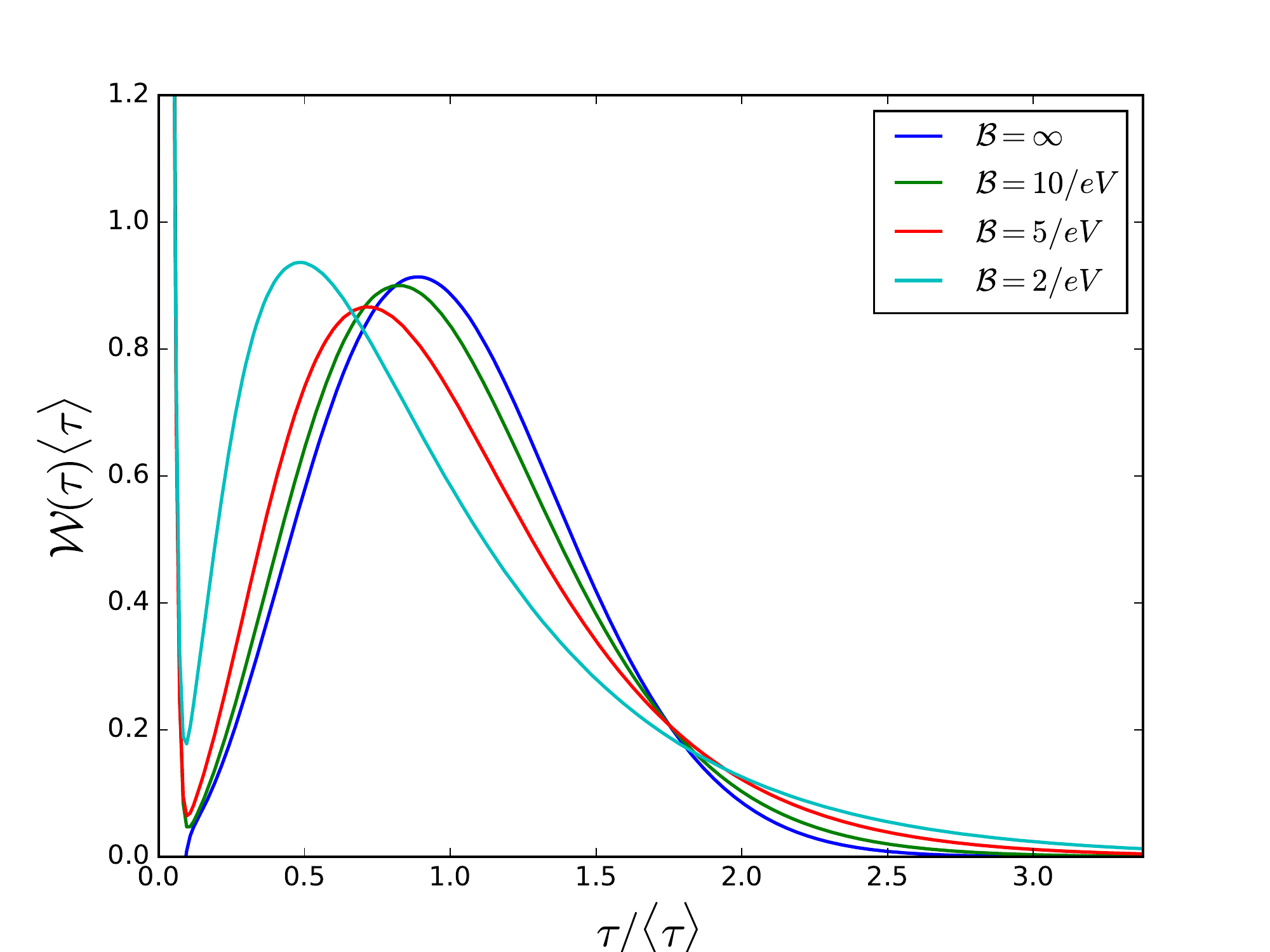}
  \caption{WTDs obtained with a smooth transmission profile. The sharpness of the transmission in Eq.~(\ref{eq:qpctransmissionfunction}) is determined by the parameter $\mathcal{B}$. We consider here the WTD for a single edge channel with an applied voltage $V$ and have used $\delta V_g = 30V$.}
  \label{fig:wtdbeta}
\end{figure}

So far, we have considered a waiting time clock with a sharp cut-off in the transmission. In reality, however, the cut-off might be smooth, corresponding to having a finite value of $\mathcal{B}$ in Eq.~(\ref{eq:qpctransmissionfunction}). In this case, electrons below the Fermi level can enter and leave the capacitor, and electrons above the Fermi level may reflect back on the QPC and never enter the capacitor. Figure~\ref{fig:wtdbeta} shows the WTD for a constant voltage $V$ with  different values of the cut-off parameter $\mathcal{B}$. For values of $\mathcal{B}$ that are much larger than the inverse voltage, the influence on the WTD is small compared to the ideal case with a sharp cut-off. For smaller values of $\mathcal{B}$, the shape of the distribution gets somewhat distorted. Still, a measurement of the WTD is clearly possible with a smooth QPC transmission.

\section{Measurement scheme}
\label{sec:fixedcouplings}

The electron waiting time clock relies on measuring the moment generating function $\chi(\lambda,\tau)$ for different values of the counting field $\lambda$ to obtain the idle time probability via the Fourier transformation in Eq.~(\ref{eq:fouriertrans}). To this end, it should be possible not only to turn the coupling on and off, but also to accurately change the strength of the coupling as well as measure the off-diagonal element of the spin density matrix. In principle, this is possible. However, as we will show now, a better strategy might be to couple several spins to the mesoscopic capacitor.

We start by considering just a single spin coupled to the capacitor during the time $\tau$. The coupling strength is denoted as $\lambda_1$. The spin is initialized in the pure state
\begin{equation}
\label{eq:initial_spin}
|\Psi\rangle=\frac{1}{\sqrt{2}}(|\uparrow\rangle+|\downarrow\rangle)
\end{equation}
 with the corresponding density matrix
\begin{equation}
  \label{eq:qubitinitial}
  \hat{\rho}_0^{(1)} = |\Psi\rangle\!\langle\Psi|=\frac{1}{2} \begin{pmatrix}
    1 & 1 \\
    1 & 1
  \end{pmatrix}.
\end{equation}
After the coupling is turned off, the density matrix reads
\begin{equation}
  \label{eq:qubitaftermeasurement}
  \hat{\rho}^{(1)} = \frac{1}{2} \begin{pmatrix}
    1 & \chi^*(\lambda_1, \tau) \\
    \chi(\lambda_1, \tau) & 1
  \end{pmatrix}.
\end{equation}
Since the coupling is fixed we cannot extract the moment generating function. However, we can calculate the probabilities of
the individual precession angles of the spin. In particular, the probability that the spin is in its initial state after the coupling has been switched off reads
\begin{equation}
  \label{eq:spinnotmoved}
  \Pi^{(1)}(\tau) = \operatorname{tr}[\hat{\rho}_0^{(1)} \hat{\rho}^{(1)}] = \frac{1}{2} \left[1 + \operatorname{Re}
  \chi(\lambda_1, \tau) \right].
\end{equation}
For $\lambda_1=\pi$, this is a crude approximation of the integral in Eq.~\eqref{eq:fouriertrans}. To improve the approximation, we couple a second spin to the capacitor.  The coupling strength of this spin is denoted as $\lambda_2$. Both spins are initially in the state given by Eq.~(\ref{eq:initial_spin}). If the couplings have been switched on during a time interval of length~$\tau$, the elements of the density matrix of the spins become
\begin{equation}
  \label{eq:jointdmelements}
  [\hat{\rho}^{(2)}]_{ij,kl} = \frac{1}{4} \chi^{(2)} \left( (i-j)\lambda_1, (k-l)\lambda_2, \tau \right).
\end{equation}
Here, the indices $i,j=0,1$ ($k,l$) refer to the first (second) spin and $\chi^{(2)}(\lambda_1,\lambda_2,\tau)$ is a joint moment generating function obtained from Eq.~\eqref{eq:mgf} by including the additional scattering phases due to the second spin. If the two spins are directly attached
to the capacitor one after another, we find that the joint moment generating function can be expressed as
\begin{equation}
\chi^{(2)} (\lambda_1, \lambda_2, \tau )= \chi (\lambda_1+\lambda_2, \tau )
\end{equation}
in terms of the moment generating function $\chi(\lambda,\tau)$ corresponding to a single spin. Calculating the probability that the spins are in their initial states after the couplings have been switched off, we find
\begin{equation}
  \Pi^{(2)}(\tau) = \frac{1}{4} \left[1 + \operatorname{Re} \left\{ \chi(\pi/2,\tau) + \chi(\pi,\tau)+\chi(3\pi/2,\tau)\right\}\right],
\end{equation}
taking $\lambda_1=\pi/2$ and $\lambda_2=\pi$. This is now a four-point approximation of the integral in Eq.~\eqref{eq:fouriertrans}. Following this line of thoughts, one can extend the idea to three or more spins, and thereby further improve the approximation of the idle time probability. For example with 3 spins with couplings $\lambda_1=\pi/3$, $\lambda_2=2\pi/3$, and $\lambda_1=\pi$, one obtains a six-point approximation of the integral.

\begin{figure}
  \centering
  \includegraphics[width=\columnwidth]{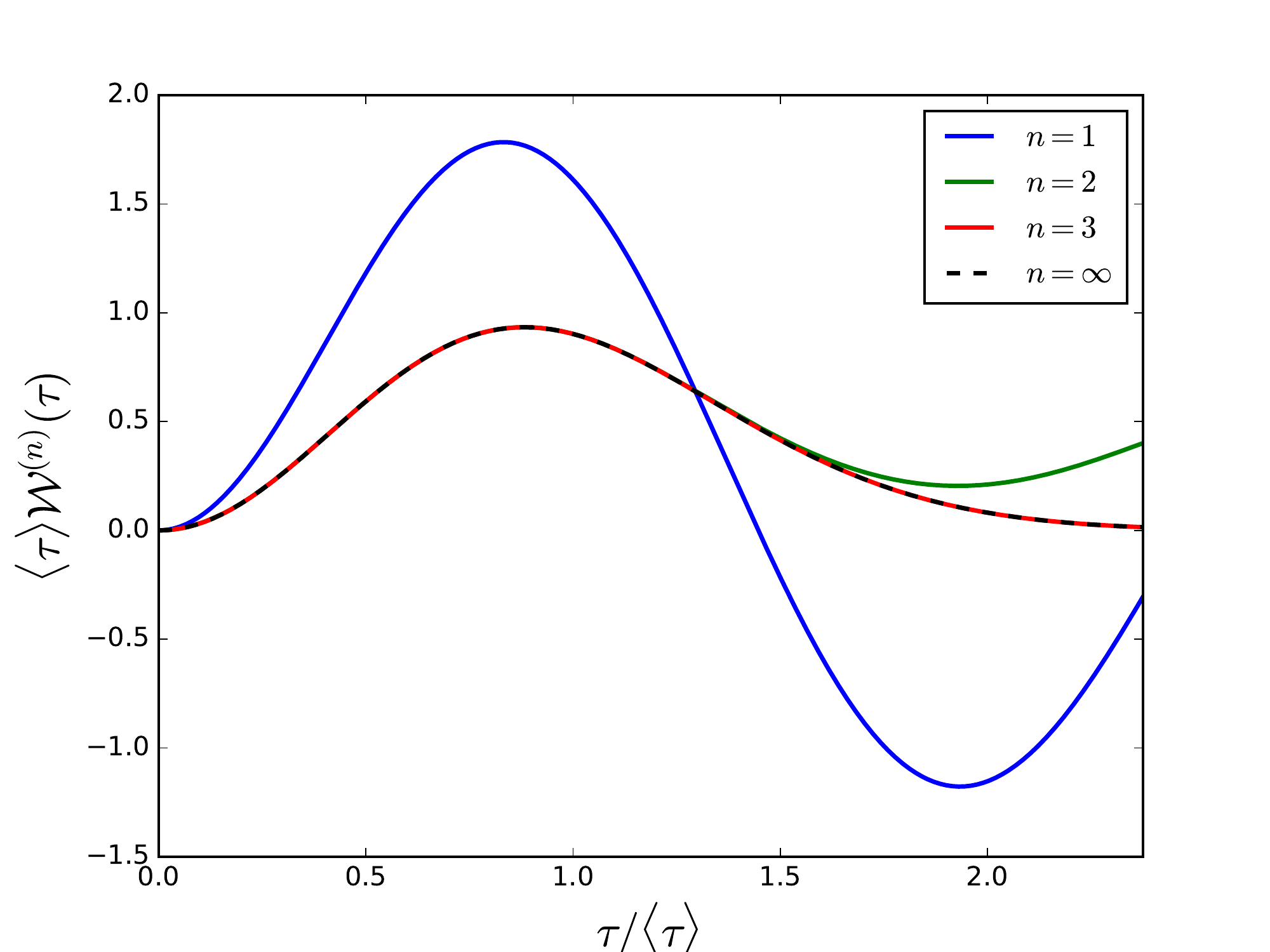}
  \caption{WTDs obtained with $n$ spins coupled to the capacitor. We consider here the WTD for a single edge channel with an applied voltage $V$ and have used $\delta V_g \gg V$. The WTD obtained with a perfect detector is shown with a dashed line.}
  \label{fig:fixedcouplings}
\end{figure}

In Fig.~\ref{fig:fixedcouplings} we show WTDs based on idle time probabilities $\Pi^{(n)}(\tau)$ measured with $n$ $(=1,2,3)$ spins. With just one spin, the WTD is only qualitatively correct for very short waiting times compared with the mean waiting time. At longer times, the WTD turns negative which is clearly not correct. However, already with two spins coupled to the capacitor, the results are much closer to the WTD obtained with a perfect detector. Only for  long waiting times, deviations become visible. With three spins coupled to the capacitor, we find essentially perfect agreement with the expected WTD for the range of waiting times shown in Fig.~\ref{fig:fixedcouplings}.

\section{Conclusions}
\label{sec:conclusions}

We have presented a quantum theory of a waiting time clock which can measure the distribution of waiting times between electrons above the Fermi sea in a mesoscopic conductor. This is an important element which so far has been missing in theories of electron waiting times. Our waiting time clock consists of a mesoscopic capacitor coupled to a quantum two-level system whose coherent precession is measured. We have demonstrated explicitly that the waiting time clock under ideal operating conditions recovers the predictions of earlier theories without a detector. We have also investigated the influence of imperfect operating conditions with two specific applications. With these advances, theories of electron waiting times can now be discussed based on a specific detector.

Our work leaves a number of questions for future investigations. The waiting time clock presented here may not be the only one that can measure the distribution of waiting times between electrons above the Fermi sea. It would be interesting to devise alternative implementations of such waiting time clocks. It might also be interesting to investigate waiting time clocks that are sensitive to correlations between waiting times, or to the electrons below the Fermi level. The distribution of electron waiting times between electrons in the Fermi sea constitutes a line of research which has not yet been addressed. Finally, the ideas presented here may form the basis for future investigations of the influence of interactions on the distribution of electron waiting times.

\begin{acknowledgments}
   We thank W.~Belzig, P.~P.~Hofer, G.~B.~Lesovik, and E.~V.~Sukhorukov  for stimulating discussions. We acknowledge the computational resources provided by the Aalto Science-IT project and by the Baobab cluster at University of Geneva. D.~D.~gratefully acknowledges the hospitality of Aalto University. The work was supported by Swiss NSF and Academy of Finland.
\end{acknowledgments}

\appendix

\section{Full scattering matrix}
\label{sec:appA}
If electrons can complete several loops inside the capacitor, the moment generating function reads
\begin{align}
  \label{eq:mgfgeneral}
  \chi(\lambda,\tau) = \det \Big( &\mathcal{S}_\text{sys}^\dagger \left(P_R + P_T
  \mathcal{S}_{-\lambda}^{(l)\dagger} P_T + P_T \mathcal{M}_{-\lambda}^\dagger P_T
  \right) \nonumber\\
  \times &\left(P_R + P_T \mathcal{S}_\lambda^{(l)} P_T +
  P_T \mathcal{M}_\lambda P_T \right) \mathcal{S}_\text{sys} \Big),
\end{align}
having introduced the matrix
\begin{equation}
  \label{eq:Mmatrix}
  \mathcal{M}_\lambda = \mathcal{S}_\lambda^{(l)}  P_R \left(1 - P_R
    \mathcal{S}^{(l)}_\lambda P_R \right)^{-1}P_R  \mathcal{S}_\lambda^{(l)}
\end{equation}
which describes processes where electrons complete more than one loop. By further manipulations, the moment generating function can be brought on the form
\begin{align}
  \label{eq:mgfgeneral2}
  &\chi(\lambda,\tau) = \det \Big( 1 + \mathcal{S}_\text{sys}^\dagger P_T \Big[ \left( \mathcal{K}_\tau^\dagger + \mathcal{K}_\tau \right) \left(e^{i\lambda/2} - 1
                 \right) \nonumber\\
  &\quad + \mathcal{K}_\tau^\dagger P_T \mathcal{K}_\tau \left(e^{i\lambda/2} - 1 \right)^2+\mathcal{R}_\tau^\lambda
                 \Big] P_T \mathcal{S}_\text{sys} \Big)
\end{align}
with
\begin{align}
  \label{eq:loopcorrections}
  \mathcal{R}_\tau^\lambda = &\mathcal{M}_{-\lambda}^\dagger P_T \mathcal{M}_\lambda + \left(\mathcal{L}^\dagger \mathcal{M}_\lambda +
    \mathcal{M}_{-\lambda}^\dagger \mathcal{L} \right) \nonumber\\
  + &\mathcal{L}^\dagger \mathcal{K}_\tau^{-\lambda \dagger} P_T \mathcal{M}_\lambda + \mathcal{M}_{-\lambda}^\dagger P_T
      \mathcal{K}_\tau^\lambda \mathcal{L},
\end{align}
where the matrix elements of $\mathcal{L}$ read
\begin{equation}
  \label{eq:Lmatrix}
  [\mathcal{L}]_{E,E'} = e^{i(E'+e\delta V_g)\tau_D} \delta(E-E'-e\delta V_g).
\end{equation}
In this case, the function in Eq.~\eqref{eq:mgfgeneral2} contains terms that are proportional to $\exp(i\lambda / 2)$. This is due to the  commutator $[P_T, \mathcal{K}_\tau]$ being non-zero.

\section{Lorentzian switching}
\label{sec:lorentzianswitching}

As an interesting aside, we consider a smooth coupling to the spin. Specifically, we take $\lambda(t)$ to be the integral of a
lorentzian,
\begin{equation}
  \label{eq:lambdalorentzian}
  \lambda(t) =  \int_{-\infty}^t \frac{-2 \tau}{t'^2+\tau^2/4} \mathrm{d} t'= -2\pi - 4\arctan(2t/\tau),
\end{equation}
such that
\begin{equation}
  \label{eq:eilambdalorentzian}
 [U_\tau^{\text{lor}}]_{t,t'}= e^{i \lambda(t)/2}\delta(t-t') = \frac{t+i\tau/2}{t-i\tau/2}\delta(t-t').
\end{equation}
It should be noted that $\lambda(t)\simeq 0$ for $t< -\tau/2$ and $\lambda(t)\simeq -4 \pi$ for $t> \tau/2$. However, due to the $4\pi$-periodicity in Eq.~(\ref{eq:eilambdalorentzian}), the value $\lambda=-4 \pi$ is equivalent to $\lambda=0$. Thus, one may think of the coupling in Eq.~(\ref{eq:lambdalorentzian}) as being non-zero only during the time interval $[-\tau/2,\tau/2]$.

In the energy representation, the elements of $U_\tau^{\text{lor}}$ are
\begin{equation}
  \label{eq:Ulor}
  [U_\tau^{\text{lor}}]_{E,E'} = \delta(E-E') - K^{\text{lor}}_\tau(E-E'),
\end{equation}
where we have defined the exponential kernel
\begin{equation}
  \label{eq:expkernel}
  K^{\text{lor}}_\tau(E) = \tau e^{- \tau E / 2} \Theta(E).
\end{equation}
Unlike the sine kernel in Eq.~(\ref{eq:sinekernel}), this kernel is only non-zero for positive energies. Thus, electrons can only absorb energy by interacting with the spin and are thus not scattered into the Fermi sea.

\begin{figure}
  \centering
  \includegraphics[width=\columnwidth]{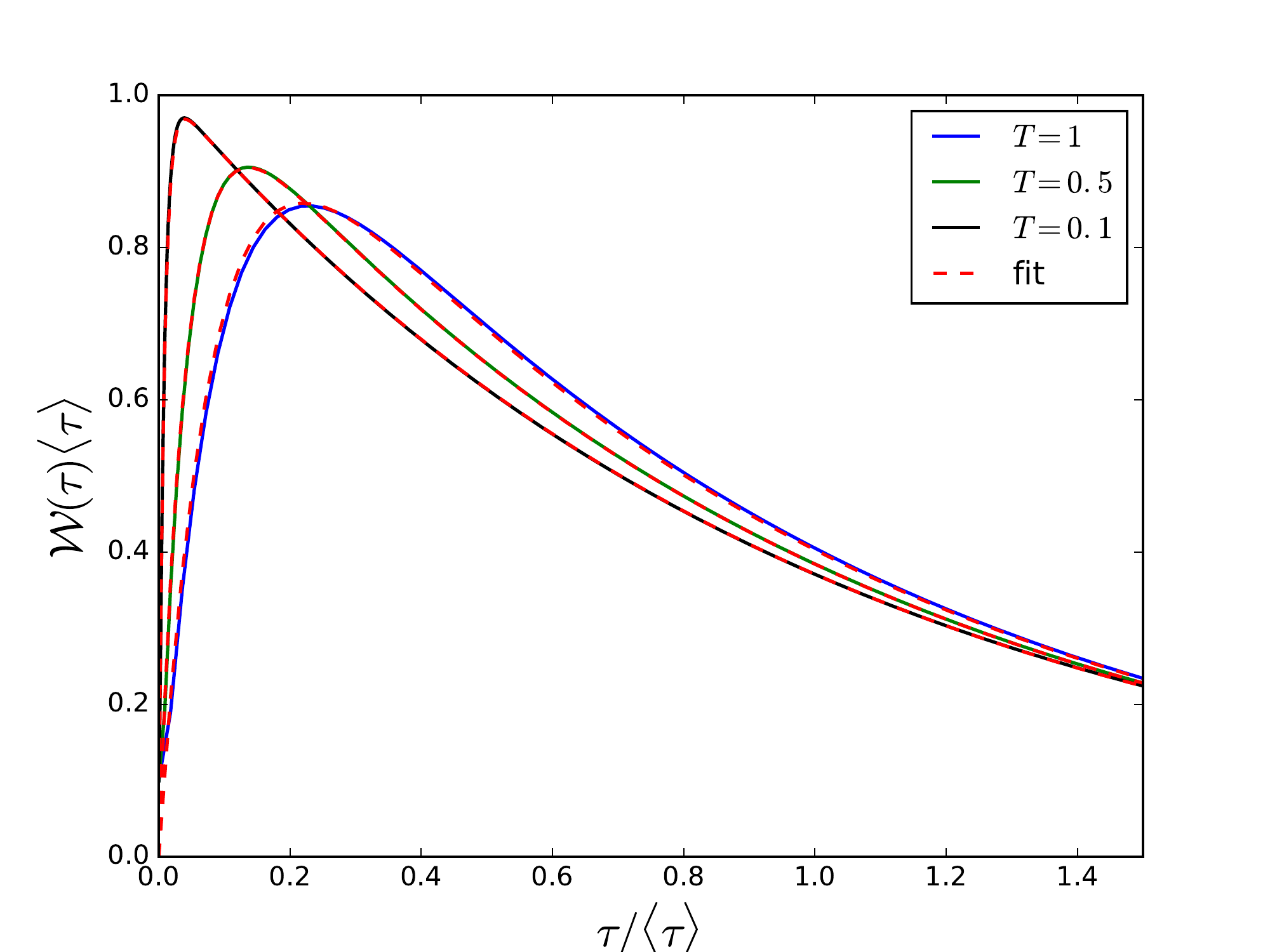}
  \caption{WTDs with a lorentzian switching of the coupling. We show results for a QPC with different transmission probabilities $T$. The dashed curves are based on Eqs.~(\ref{eq:surmise}, \ref{eq:WTDsurmise}).}
  \label{fig:lorentzianwtd}
\end{figure}

Nest, we evaluate Eq.~\eqref{eq:mgfchiral} and find
\begin{equation}
  \label{eq:itplorentzian}
  \chi(\lambda) = \det \left(1 -  Q^\text{lor}_\tau
  \right)
\end{equation}
with
\begin{equation}
  \label{eq:Qlor}
  Q^\text{lor}_\tau =  \mathcal{S}_\text{sys}^\dagger P_T \mathcal{K}_\tau^\text{lor}    P_T \mathcal{S}_\text{sys}
\end{equation}
and
\begin{equation}
\label{eq:expkernel2}
  [\mathcal{K}_\tau^\text{lor}]_{E,E'} = \tau e^{-|E-E'|\tau/2}.
\end{equation}
Surprisingly, by comparing these expressions with Eqs.~(\ref{eq:Qmatrix},\ref{eq:itponlypositive}), we see that Eq.~(\ref{eq:itplorentzian}) takes the form of an idle time probability, however, with the kernel given by Eq.~(\ref{eq:expkernel2}). Thus, without further justification, we consider in the following $\chi(\lambda)$ as the idle time probability and evaluate the corresponding WTD by differentiating it twice with respect to $\tau$.

In Fig.~\ref{fig:lorentzianwtd} we show WTDs for a QPC with transmission $T$ obtained in this way. The mean waiting time is still given by Eq.~(\ref{eq:qpcmeanwait}), however, the WTDs are different from those in Fig.~\ref{fig:wtd}b. The WTD appears to depend linearly on $\tau$ at short times and eventually decays exponentially at long times. This resembles the WTD for a resonant level in the high-bias limit\cite{brandes08}
\begin{equation}
\label{eq:surmise}
\mathcal{W}(\tau)=\frac{\Gamma_L\Gamma_R}{\Gamma_R-\Gamma_L}(e^{-\Gamma_L\tau}-e^{-\Gamma_R\tau}),
\end{equation}
where $\Gamma_L$ and $\Gamma_R$ are the rates at which electrons enter and leave the level. The mean waiting time reads
\begin{equation}
\langle\tau\rangle=\frac{\Gamma_L+\Gamma_R}{\Gamma_L\Gamma_R}.
\label{eq:WTDsurmise}
\end{equation}
Based on the similarity, we surmise that Eq.~(\ref{eq:surmise}) also describes the WTDs in Fig.~\ref{fig:lorentzianwtd}. The rate $\Gamma_R$ can be determined from the mean waiting time. We then use $\Gamma_L$ to fit our results for full transmission and find excellent agreement. For the results with finite transmission, we keep $\Gamma_L$ fixed and extract $\Gamma_R$ from the mean waiting time which depends on the transmission. With this approach, we can fully account for all results in Fig.~\ref{fig:lorentzianwtd}. Further investigations of these findings are left for future work.

\end{document}